\documentclass[9pt,twocolumn,twoside]{osajnl}

\usepackage{graphicx,amsmath,amssymb,color}
\journal{Arxiv_ol}

\setboolean{shortarticle}{true}

\newcommand{\redline}[1]{ {\color{black}{#1}}}  
\newcommand{\strikeout}[1]{ }

\title{Refractive-Index-based ultrasound sensing with photonic crystal slabs}

\author[1,2,*]{Eric Y. Zhu}
\author[1,2]{Cory Rewcastle}
\author[1,2]{Raanan Gad}
\author[1]{Li Qian}
\author[1,2,$\dagger$]{Ofer Levi}

\affil[1]{Department of Electrical and Computer Engineering, University of Toronto, 10 Kings College Road, Toronto, Ontario M5S 3G4, Canada}
\affil[2]{Institute of Biomaterials and Biomedical Engineering, University of Toronto, 164 College Street, Toronto, Ontario M5S 3G9, Canada}

\affil[*]{eric.zhu@utoronto.ca}
\affil[$\dagger$]{ofer.levi@utoronto.ca}

\begin{abstract}

We demonstrate {ultrasound detection} with 500-$\mu\mathrm{{m}}$-diameter photonic-crystal slab (PCS) sensors fabricated from CMOS-compatible technology.  
An ultrasound signal impinging a PCS sensor causes a local modulation of the refractive index (RI) of the medium (water) in which the PCS is immersed, resulting 
in a periodic spectral shift of the optical resonance of the PCS.  
The acoustic sensitivity 
is found to scale with the index sensitivity $S$ and quality factor $Q$.
A noise equivalent pressure (NEP) of 650 Pa 
{with averaging (7.4 Pa$/\sqrt{\mathrm{Hz}}$)} 
{and relative wavelength shifts of up to 4.3$\times10^{-5}$ MPa$^{-1}$}
are measured.  The frequency response of the sensors is  observed to be flat from 1-20 MHz, with the range 
limited only by our measurement apparatus.  
\end{abstract}

\begin{document}

\maketitle


The ability to measure ultrasound has many applications, ranging from biomedical imaging to non-destructive testing.  
The common workhorse for such applications is the piezoelectric transducer \cite{cobbold2006foundations}, which uses the piezoelectric effect to both transmit and receive ultrasound signals. 
However, these devices suffer from narrow bandwidths, and poor sensitivity. 
Recently, work on capacitive micromachined ultrasound transducers (cMUTs), which are fabricated using CMOS-compatible technology, has helped to improve signal detection bandwidths, 
but the detection and generation of ultrasonic signals nevertheless are still performed electrically. When placed into an array, both piezoelectric and cMUT devices experience significant electrical crosstalk \cite{bayram2007finite}  as well as electromagnetic interference (EMI) in general.  Additionally, commercially available piezo electric arrays, due to miniaturization, have compromised sensitivity characteristics compared to their larger counterparts.  

One solution to eliminating cross-talk and EMI is to detect the acoustic signals optically \cite{dong2017optical,wissmeyer2018looking}. Such implementations often couple the optical and acoustic signals using an optical cavity or waveguide. 
%
\strikeout{
The use of an optical cavity can also help with miniaturization of the ultrasound sensor; given that the optical wavelengths used for interrogation  are typically ( $\approx$ 1 $\mu$m) much shorter than the acoustic wavelengths measured, the ability of an optical field to interact with the acoustic signal over many round trips inside a resonator  can result in much greater sensitivity.  
}
The development of  optical ultrasound sensors has been motivated primarily by their use in photoacoustics, a biomedical imaging  modality \cite{wang2016practical} that combines the benefits of both ultrasound and optical imaging.  
\strikeout{It involves the delivery of high intensity laser pulses with mJ pulse energies 
into tissue that result in their conversion into 
broadband ultrasound waves.  
With the tissue essentially acting as an ultrasound 'source', the medium can be imaged in a way that would otherwise be inaccessible with conventional optical means.  
However, even with such large pulse energies, ultrasound signals can still be significantly attenuated after travelling through tissue, 
in some cases requiring sensors with detection levels below 100 Pa \cite{rosenthal2014sensitive, yao2014sensitivity} in order to be detected.}
This technique often requires high acoustic sensitivity,
in some cases requiring sensors with detection levels below 100 Pa \cite{rosenthal2014sensitive, yao2014sensitivity}.
As with conventional ultrasound imaging, an array of sensors is also required to form an image, necessitating the miniaturization of each sensor.  
The requirements of small size,  high sensitivity, and  large bandwidth cannot be satisfied by current piezo-electric or CMUT transducers,  but they can be met with optically-based sensors.  

All-optical ultrasound sensors have been implemented in bulk Fabry-Perot resonators \cite{zhang2008backward}, 
fiber Bragg gratings (FBG) \cite{rosenthal2014sensitive}, and fiber-tip cavities \cite{guggenheim2017ultrasensitive}, with sensitivities nearing 10 Pa 
and bandwidths exceeding well over 20 MHz.  
Recent integrated solutions include surface-plasmon resonance (SPR) devices \cite{zhou2018ultrasound} and microring resonator devices \cite{ling2011high} with 
bandwidths exceeding 100 MHz and {noise levels} below 25 Pa (respectively).
These optical sensors, however, have several drawbacks; 
their fabrication methods may lead to significant device variability \cite{zhou2018ultrasound, ling2011high,guggenheim2017ultrasensitive}, 
they may be extremely fragile \cite{zhou2018ultrasound}, 
or may be difficult to incorporate into a multi-pixel device \cite{rosenthal2014sensitive}.

CMOS-compatible fabrication can yield minimal device-to-device variation, meaning that creating an array of identical sensors is straightforward.  
{
Recent CMOS-compatible all-optical ultrasound sensors include waveguide- \cite{tsesses2017modeling} and ring-resonator-based \cite{leinders2015sensitive} devices.  
The presence of an ultrasound signal results in the 
mechanical deformation and elasto-optic change of the light-guiding medium, and
in highly mode-confined architectures such as  \cite{tsesses2017modeling} and \cite{leinders2015sensitive},
these effects result in the modulation of the the optical intensity.  
}
%

{
By contrast, the sensors we use in this work, which are also CMOS-compatible, are engineered to have significant portions (up to 22\%) \cite{el2010sensitivity}  of the field mode energy reside outside of the guiding structure, 
}
allowing us to observe the
change in the index of refraction of the water surrounding the sensor when an acoustic signal is present.
Our ultrasound sensors are miniature (500-micron diameter) photonic crystal slab (PCS) devices (previously used for index sensing \cite{el2010sensitivity}) capable of noise equivalent pressures (NEPs) of {650 Pa with averaging (7.4 Pa$/\sqrt{\mathrm{Hz}}$)},   acoustic bandwidths extending from 1 MHz to 20 MHz, and 
relative wavelength shifts of up to 4.3$\times10^{-5}$ MPa$^{-1}$.  
%

In what follows, we will investigate the physics behind the ultrasound sensing, 
and provide a path toward fabricating the next generation of small-footprint 
PCS sensors that will have sensitivities rivaling the current state-of-the-art.  

\begin{figure}[t!]
	\centering
	\includegraphics[width=8cm]{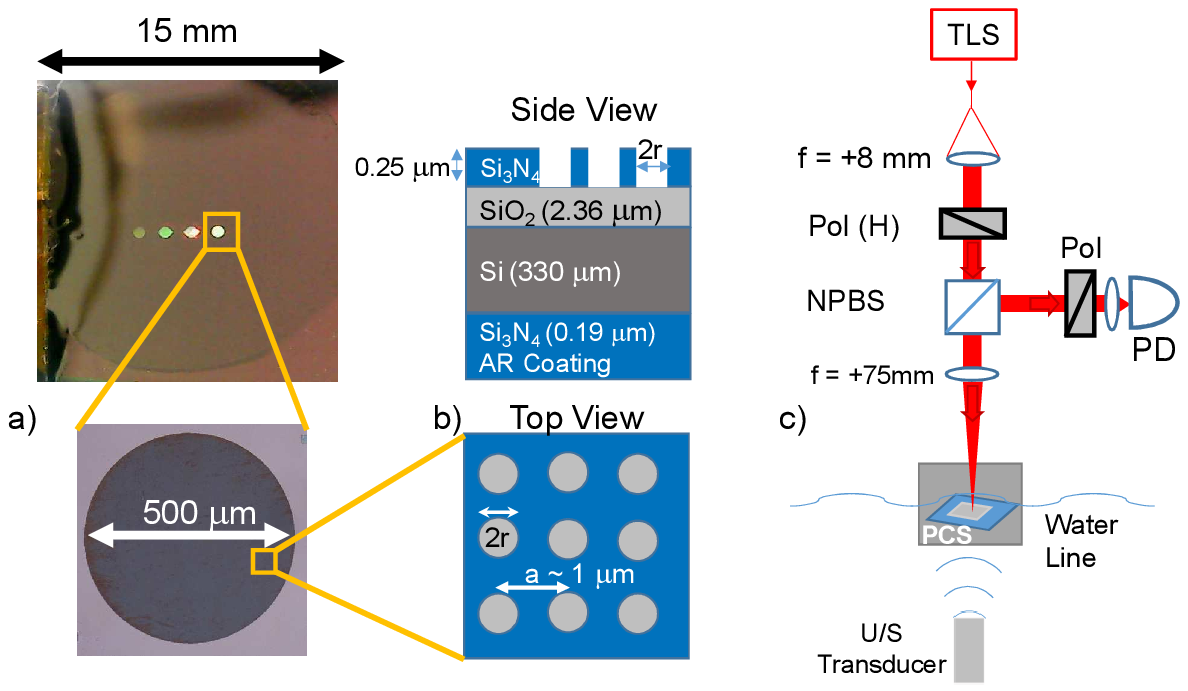}
	\caption{\label{fig:Chip_Setup}
	a)  A photograph of the 15 mm $\times$ 15 mm chip containing four photonic crystal slab (PCS) sensors.  
	A 10x optical microscope image reveals a 500-micron diameter region patterned with a dense array of nanoholes.  
	b)  A schematic showing that the nanoholes are arranged in a simple square lattice whose period $a$ is approximately 1 micron; the 
	hole radius $r$ (held constant for each sensor) varies from 100 nm to 350 nm.  A cross-section of the chip (not to scale) reveals that the PCS grating is written on a top layer of 
	stoichiometric silicon nitride (Si$_3$N$_4$), which sits on top of a layer of silicon dioxide (SiO$_2$) and a thick silicon (Si) substrate.  
	c)  
{	
	Experimental setup.  	
	Tunable laser source (TLS), linear polarizer (Pol), non-polarizing beam splitter (NPBS), 
	f=+8 mm out-coupling and f= +75 mm focussing lenses, and photodiode (PD).  
	}
\strikeout{	
		When used as an ultrasound (U/S) sensor, the front (PCS) side of the chip is immersed in water; the AR-coated backside is interrogated with a tunable laser source (TLS) that
	has been focused down to a 90 $\mu{\mathrm{m}}$ spot with the help of an $f = 8$ mm asphere and $f = 75$ mm planoconvex lens.  
	A non-polarizing beam splitter (NPBS) re-directs the backreflected beam into a fast photodiode (PD).  
	A polarizer (Pol) ensures only linearly-polarized light is launched into the sensor.  
	}
	}
\end{figure}
Our PCS devices are fabricated on a 15 mm $\times$  15 mm silicon (Si) wafer with silicon dioxide (SiO$_2$) 
and stoichiometric silicon nitride (Si$_3$N$_4$) layers (Fig. \ref{fig:Chip_Setup}).  
An optical microscope image of a single PCS is shown in Fig. \ref{fig:Chip_Setup}a.  
The PCS consists of a 500-$\mu\mathrm{{m}}$ diameter region of silicon nitride that has been patterned with a periodic array of nanoholes through electron beam lithography. 
{The PCS optical resonance exhibits an asymmetric lineshape (Fig. 2, red trace) known as the Fano resonance; 
the asymmetry arises due to the interplay between the Fabry-Perot resonance of the thin Si$_3$N$_4$ layer and the guided resonance of the PCS structure \cite{SFan2002}.
}
The center wavelength $\lambda_0$ of the optical resonance is a function of the 
index of refraction $n$ of the fluid  in which the PCS is immersed.  
We define the index sensitivity $S$ as the change in $\lambda_0$ as $n$ varies:  $S = d{\lambda_0}/dn$.
The lattice constant $a$ is chosen to be approximately 1.0 micron to allow for the optical resonance to reside spectrally within the 1550 nm communications band.  
The nanohole diameter for each PCS device on the chip is different to provide us with sensors of varying linewidths (quality factors $Q$) and index sensitivities $S$.  
The various layers of the wafer/device are shown in Fig. \ref{fig:Chip_Setup}b, with a thin bottom layer of Si$_3$N$_4$ 
acting as an anti-reflection (AR) coating to allow for interrogation of the PCS from the backside with 1550 nm light. 

The front (PCS) side of the sensor is immersed in distilled water (Fig. \ref{fig:Chip_Setup}c), and normal to incoming acoustic waves produced by a precalibrated piezoelectric ultrasound transducer (UST);
{
the calibration was performed with an Onda HMB-200 hydrophone, which has a flat frequency response from 1-40 MHz.
}\hspace{-3.5pt}
The PCS sensor is held in place with a 
custom holder. 
A linearly-polarized continuous-wave (CW) optical beam from a narrow linewidth {(< 100 kHz)} tunable laser source (TLS, {Keysight 81960A}) interrogates the PCS from the back side {with incident power typically around 10 mW}; the spot size of this beam is focused down to  90 microns, significantly smaller than the 500-micron sensor size.  
The back-reflected light {($\sim$ 2 mW)} from this interrogating beam is directed to a fast photodiode (PD in Fig. 
\ref{fig:Chip_Setup}c) using a non-polarizing beam splitter (NPBS).  
{
A cross-polarizer (Pol) placed immediately before the PD removes Fabry-Perot fringes from the silicon substrate; it also reduces the optical power incident on the PD
 to 20 $\mu$W, well below the detector's saturation (50 $\mu$W).   
}
The PD is a New Focus 1811 detector; 
its electrical output is split with a bias-tee (not shown in Fig. \ref{fig:Chip_Setup}c) into a DC component used to measure the reflectance of the sample, 
and an (amplified) AC ($>$ 70 kHz) component  used for ultrasound sensing.  Both signals are digitized with a 300 MHz bandwidth real-time  oscilloscope.  

\strikeout{
To measure the acoustic sensitivity of our devices, a piezoelectric ultrasound transducer (UST) is placed below the PCS in a water tank filled with distilled water.  
}
\strikeout{
The distance between the PCS and UST is adjusted depending on the acoustic frequency excited and size of the UST element; 
the distance is large enough so that the device measures the properties of the acoustic beam outside of the Fresnel zone, 
but short enough so that the acoustic beam diameter remains small (typically $<$ 6 mm).  }
\strikeout{The UST positioning in all three dimensions is optimized with the use of a translation stage.  
All USTs used in the experiment are calibrated beforehand with an Onda HMB-200 hydrophone, which has a flat frequency response from 1-40 MHz.
}

\begin{figure}[hb!]
\centering
	\includegraphics[width=8.5cm]{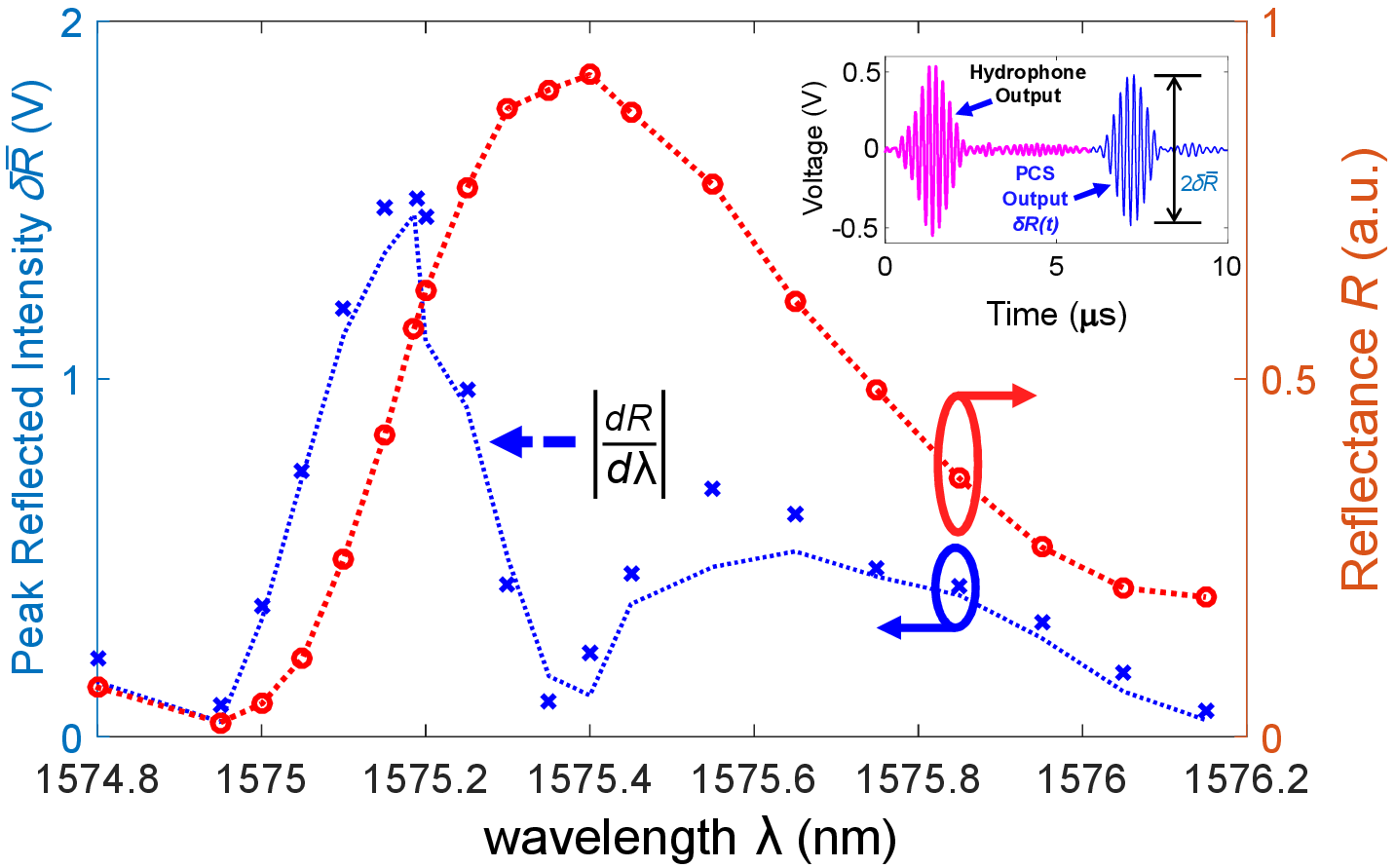}
	\caption{ \label{fig:SensVsWL}
	The red curve shows the reflectance $R$ of the PCS as a function of wavelength.  The asymmetric lineshape is called a Fano resonance.  
	The dotted blue curve is the absolute value of the reflectance slope $|dR/d\lambda|$; it matches well with the {peak PCS reflected intensity $\delta{\overline{R}}$} ($\times$ data points, {defined in inset}).  
	\strikeout{This plot helps us choose the optimum wavelength ($\lambda = $ 1575.15 nm) at which to interrogate the PCS.}  
	Inset shows the temporal response of a sensor to an impinging acoustic signal composed of a Gaussian envelope (FWHM = 1 $\mu$s) with a 5 MHz carrier; {
	the blue trace gives a typical PCS sensor reflectivity change ($\delta{R(t)}$, Eqn. 1), 
	while the pink trace provides the hydrophone output $\delta{V(t)}$.  
	The curves were shifted in time and re-scaled in voltage to improve legibility.  }
%
	}
\end{figure}

We now elaborate further on the sensing mechanism for our PCS device.  An ultrasound pulse is essentially a time-varying pressure change $\delta{P(t)}$ in the water medium; 
{such a pressure change results in a modulation $\delta{n_w}$ of the local index of refraction of water}, $\delta{n_w}=\frac{d n_w}{dP}\times\delta{P(t)}$; 
the value of $\frac{dn_w}{dP}$ has been measured to be $1.38\times10^{-4}$ refractive index units per MPa (RIU/MPa) \cite{waxler1963effect}.     
This change $\delta{n_w}$ causes the optical resonance of the PCS to shift spectrally, which is dictated by the index sensitivity $S\equiv \frac{d\lambda_0}{dn}$ of the sensor.  
When the interrogating optical wavelength is close to the optical resonance, the index modulation $\delta{n_w}$ (and consequently, 
the ultrasound pulse) is mapped onto the reflected optical intensity $\delta{R(t)}$, 
{which is obtained experimentally from the AC component of the PD output (Fig. \ref{fig:Chip_Setup}c)}.  
Put together, we obtain the expression:
\begin{equation}
	\delta R(t)=(1+r)\times \delta{P(t)}\times \frac{dn_w}{dP} \times \frac{d\lambda_0}{ dn} \times  \frac{dR}{ d\lambda}.
	\label{eq:IntensityVsRefl}
\end{equation}
We note that Eqn \ref{eq:IntensityVsRefl} contains a pressure enhancement factor $(1+r)$ due to the water-PCS interface \cite{alcock2002sensitivity, cobbold2006foundations}, where $r$ ($= 0.84$) is the acoustic Fresnel reflection coefficient for water to silicon.  The final term in Eqn. \ref{eq:IntensityVsRefl}, $\frac{dR}{d\lambda}$, is the reflectance slope of the PCS optical resonance (Fig. \ref{fig:SensVsWL}, red trace).  
{Neither the mechanical deformation nor the 
elasto-optic changes in the PCS device that result from an impinging acoustic 
pressure contribute substantially to the optical 
resonance shift; both effects were evaluated and found to be more than an order of magnitude smaller than the water-index mechanism 
(Eqn. \ref{eq:IntensityVsRefl}). 
This is due to the relatively low elasto-optic coefficient for silicon nitride \cite{capelle2017polarimetric}, and the thick substrate  (> 300 $\mu$m) (Fig. \ref{fig:Chip_Setup}b) upon which the PCS is fabricated.
}

For a given resonance, the slope $\frac{dR}{d\lambda}$ varies depending on the wavelength of the interrogating laser, and Eqn 
\ref{eq:IntensityVsRefl} implies that the sharper the slope, the greater the acoustic sensitivity.  Figure \ref{fig:SensVsWL} demonstrates just that; 
where the optical reflectance (red plot) slope is greater, the peak reflected optical intensity $\delta{\overline{R}}$ (blue data points) is also commensurately larger.  The dotted blue 
line is the derivative $dR/d\lambda$, and agrees well with the acoustic sensitivity to within a multiplicative factor.  
\strikeout{
The acoustic signal for this 
experiment is provided by an Olympus immersion transducer (C326-SU) that
generates a peak pressure of 156 kPa at the water-PCS interface.  }
The UST providing the acoustic signal is driven by a  {transform-limited} 
Gaussian pulse (full width at half maximum, FWHM = 1 $\mu$s)  with a 5 MHz carrier frequency {(and FWHM bandwidth of 0.44 MHz)}.  
The inset of Fig. 2  shows the reflected optical intensity $\delta{R(t)}$ (blue trace) from the PCS in response to an impinging  ultrasound pulse.  
{A slight ringing following the Gaussian pulse (associated with the transducer) 
is observed with both the PCS sensor and a reference hydrophone (pink trace).  }
{The peak optical reflected intensity $\delta{\overline{R}}$ is measured by fitting the trace  $\delta{R(t)}$ to
a Gaussian waveform}.  In order to overcome the intensity noise of the 
laser source at 1-20 MHz, this final trace is an average of 64 real-time traces.   The properties of the PCS used for this measurement are given in 
Table \ref{tab:Sensors}, in the column labeled 'High Q'.   

{
The maximal slope $\frac{dR}{d\lambda}$ correlates well with the optical resonance quality factor $Q$.  
By replacing $\frac{dR}{d\lambda}$  with $Q$ in Equation \ref{eq:IntensityVsRefl}, a design rule emerges.  At a given 
wavelength, the PCS with the larger $S\times{Q}$ product will have the higher acoustic sensitivity.  }

\begin{figure}
	\centering
	\includegraphics[width=8cm]{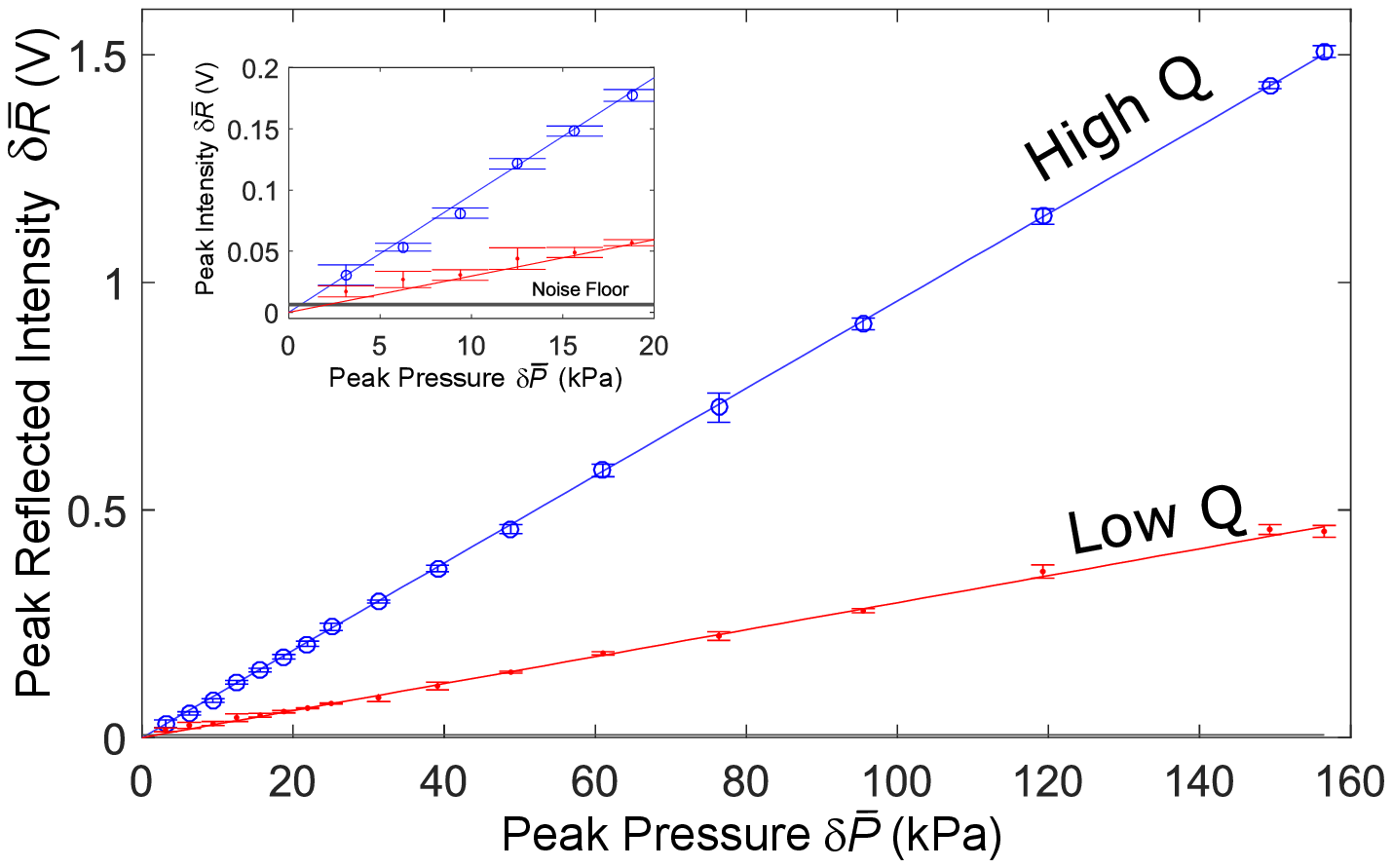}
	\caption{ 	\label{fig:C3_vs_C4}
		The {ultrasound-modulated peak optical reflection intensity $\delta{\overline{R}}$} of two different PCS sensors are plotted as a function of the peak applied pressure $\delta{\overline{P}}$.  
		The acoustic sensitivity slope of the `High-Q' sensor  (9.6 mV/kPa) is approximately 3$\times$ stronger than that of the 'Low-Q' sensor (3.0 mV/kPa).  
\strikeout{
		This ratio corresponds to the $S\times Q$ product ratios of the two sensors (see Table\ref{tab:Sensors}).}
		The inset shows the same plot in the 0-20 kPa range; the measurement noise floor is denoted by a horizontal line. 
	}
\end{figure}

In Fig. \ref{fig:C3_vs_C4}, the {ultrasound-modulated peak optical reflection intensity $\delta{\overline{R}}$ (Eqn. 1)} of two different PCS sensors present on the same 15 mm x 15 mm wafer (Fig. \ref{fig:Chip_Setup}) are plotted as a function of the peak applied pressure.  
The optical properties of each PCS are given in Table \ref{tab:Sensors}; 
one is labeled `High $Q$', and the other `Low $Q$'.  
The experimental procedure used in obtaining Fig. \ref{fig:C3_vs_C4} is similar to what was used before; the interrogating wavelength for each sensor is optimized using the procedure from Fig. \ref{fig:SensVsWL}.  

{Figure \ref{fig:C3_vs_C4} provides us with insight into several aspects of the acoustic sensitivity of our devices.}
First, the   peak PCS  optical reflection intensity $\delta{\overline{R}}$ remains linear over the 10-160 kPa pressure range, indicating a large dynamic range.  
Secondly, the experimental acoustic sensitivities $\delta{\overline{R}}/\delta{\overline{P}}$  for each 
PCS device (Table 1) can be  obtained from the slope of each data set. 
The expected values for each device (calculated from Eqn \ref{eq:IntensityVsRefl}) are also included in Table 1 
for reference.  
{
As the experimental and expected values are within 35$\%$ of each other, 
this gives us high confidence that the ultrasound-sensing mechanism is due primarily to changes in the refractive index of water.
}

\begin{table}[b!]
\centering
\caption{\bf Optical and Acoustic  Properties of PCS Sensors \label{tab:Sensors}}

\begin{tabular}{ccc}
\hline
		Property			& 			Low-Q  			& 			High-Q  \\
\hline
Peak $\lambda_0$ (nm)	 & 			1560 				& 		1575	    \\
Linewidth $\Gamma$ (nm)	 	& 			3.1 				& 			0.71		    \\
Quality Factor $Q$ ($=\lambda_0/\Gamma$)   & 				503				& 	       2220		\\
{Slope $dR/d\lambda$ (V/nm)} &                       {45 }                         &             {303 }                      \\
$S$ {($=\frac{d\lambda_0}{dn}$, nm/RIU)}    & 			216					&		165		\\
$S\times Q$ {($\times 10^5$ nm/RIU)}		&			$1.09$		&		$3.66$ \\
\hline	
\hline
{Measured Sensitivity }$\frac{\delta{\overline{R}}}{\delta{\overline{P}}}$ (V/MPa)		&			3.0					&		9.6					\\
Expected $\frac{\delta{\overline{R}}}{\delta{\overline{P}}}$ (Eqn. 1) (V/MPa)		&			2.5				&		12.7					\\
Wavelength Shift $\frac{1}{\lambda_0}\frac{d\lambda}{dP}$ ($\times10^{-5}$ MPa$^{-1}$) & $4.3$& $2.1$\\
Noise Level (mV) 		&				6.2					&		6.2			\\
{NEP (kPa)} 	&		{2.1} 					&	{0.65}				\\			
{Normalized NEP ($\mathrm{kPa/}\sqrt{\mathrm{Hz}}$)} 	&			{0.024}&			{0.0074}				\\		
\hline
\end{tabular}
  \label{tab:shape-functions}
\end{table}

The inset of Fig. \ref{fig:C3_vs_C4}, which shows the peak  PCS reflected intensity $\delta{\overline{R}}$ for low peak pressures along with the measurement noise floor (grey horizontal line), provides us with a graphical means to gauge the noise-equivalent pressure (NEP) of each sensor.  Not surprisingly, the NEP of the `Low Q' device (2.1 kPa, {or 23.8 Pa/$\sqrt{\mathrm{Hz}}$ when normalized to the measurement bandwidth}) is worse than the `High-Q' device (0.65 kPa, or {7.4 Pa/$\sqrt{\mathrm{Hz}}$}).  
{As each data point in Fig. \ref{fig:C3_vs_C4} is obtained from an average of 64 traces, the measurement bandwidth (which is related inversely to the 
integration time) is 7.8$\times$10$^3$ Hz.}

We notice in Table 1 that the device with the higher $S\times{Q}$ product also has the correspondingly higher acoustic sensitivity.  
By simply having a design for a PCS with an $S\times Q$ product that is 10-fold greater, we would expect \cite{IanWhite:08} 
a 10-fold improvement in the acoustic sensitivity.  
Previous work \cite{Costa2013, Raanan2015} has shown such an improvement is readily achievable with only slight modifications to the device geometry, 
\redline{
which include the removal of the Si substrate and the use of non-circular (elliptical) nano-holes. 
While the optical absorption of water limits the $Q$ of these devices to $\sim 15,000$ at 1550 nm, 
higher-$Q$ devices can be achieved with shorter interrogating wavelengths 
($\sim$ 900-950 nm) where water absorption is greatly reduced. 
}

{Additionally, our PCS sensors can be further reduced in size from their 500-micron diameters, since the effective sensing region, 
determined by the spot size of the interrogating optical beam, is only 90 microns wide (1/$e^2$ diameter).  }
{Reducing the sensor to a 50-micron diameter footprint 
would allow for high-density arrays of sensors useful for imaging.  
This should be feasible using silicon photonics fabrication techniques without compromising the acoustic sensitivity or bandwidth, as the index-sensing mechanism is expected to be broadband and the PCS design rule $S\times Q$ is independent of device dimensions (so long as the PCS contains more than $30\times30$ unit cells 
\cite{grepstad2013finite}).   
}

\begin{figure}[t!]
\centering
	\includegraphics[width=8cm]{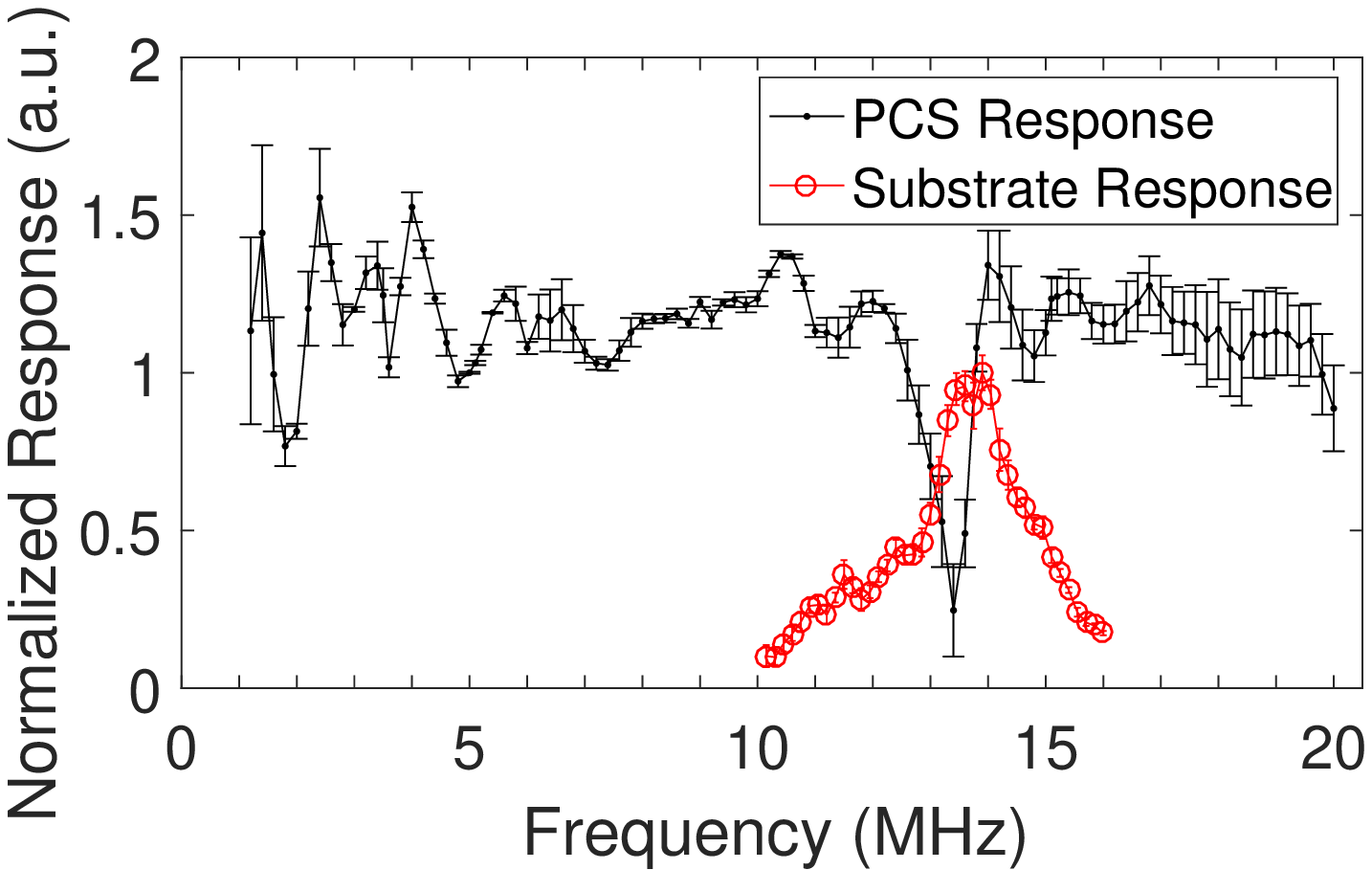}
	\caption{ \label{fig:FreqResponse}
		Frequency response for the `High-Q' PCS sensor (black line), 
		{normalized to the response at 5 MHz}.  
		The frequency was scanned from 1-20 MHz, at 0.2 MHz steps.  
		\strikeout{
		The response for the 'Low-Q' sensor is similar.  
		}
		A dip at 13.4 MHz is present due to the silicon substrate of the sensor acting as an acoustic oscillator at this frequency.  The acoustic response of the silicon substrate is shown in the red trace.  
		A flat response for the PCS is otherwise observed from 1-20 MHz.
	}
\end{figure}

The acoustic frequency response of the PCS sensor from 1-20 MHz is assessed using three different immersion transducers.  
At all frequencies, the transducers are excited with 2.5-microsecond Gaussian pulses (corresponding to a signal bandwidth of 0.18 MHz).
Figure \ref{fig:FreqResponse} gives the normalized frequency response of the high-Q PCS sensor, which shows that the acoustic sensitivity of the PCS remains relatively flat (within 30$\%$) over the 1-20 MHz range, 
with the exception of a sharp dip at 13.4 MHz.  
This dip is due to the silicon substrate (red trace, Fig. \ref{fig:FreqResponse})
of the PCS acting as a {mechanical} resonator for longitudinal acoustic waves; 
the thickness of the substrate (330 $\mu$m, Fig. \ref{fig:Chip_Setup}) corresponds roughly to half the acoustic wavelength at 13.4 MHz in silicon (c $\sim$ 8400 m/s \cite{HopcroftSiYoungsModulus}).  
{At frequencies close to this resonance, much more acoustic power is coupled into the silicon substrate, which reduces the amount of acoustic 
power at the water-PCS interface, and lowers the pressure amplitudes experienced by the water at that interface.  }
{The acoustic sensitivity of the Si substrate is assessed by optically interrogating a region of the sample without a PCS present, and 
measuring the shifts in the substrate's \emph{optical} Fabry-Perot resonance in the presence of acoustic excitation (Fig. \ref{fig:FreqResponse}, red trace).  }
Similar resonances have previously been observed in CMUT devices \cite{LadabaumIEEE2000, guldiken2009dual}, 
and can be mitigated by either thinning down the substrate or by placing a backing layer behind it \cite{jin2001characterization, hu2018stretchableSciAdv}.

While we have only been able to demonstrate a flat frequency response up to 20 MHz in this work, the limitation is due to the ultrasound sources available to us.  
We anticipate the acoustic frequency response of the PCS device to be much broader band, 
and plan to demonstrate this in future work with a broadband photo-acoustic ultrasound source.
As with other optical ultrasound sensors \cite{dong2017optical,wissmeyer2018looking}, a bandwidth of up to 100 MHz
may be possible with a smaller diameter PCS.  
Additional future work will include bundling many identical PCS sensors into an array.  
In place of free-space optical interrogation, each sensor can be fiber butt-coupled from the backside.  By adding an elastomeric backing \cite{hu2018stretchableSciAdv} that serves to also hold the fiber facets in place, 
an easily-moldable spatially-multiplexed sensor array requiring only a single interrogating light source can be realized.   


In summary, we have demonstrated a photonic crystal slab-based ultrasound sensor that has a flat frequency response up to at least 20 MHz. 
{
By engineering the slab so that a substantial amount of the field mode energy resided outside of the light-guiding medium, we were able to measure the pressure-induced changes in the refractive index of water surrounding the slab to detect ultrasound signals on the order of 1 kPa.  
}
This work also provides us with guidance on reaching greater sensitvities, simply by improving the index sensitivity $S$ and quality factor $Q$ of our PCS design. 
Additionally, the CMOS compatible nature of our sensors indicates that they can be mass-produced in such a way that device variability would be minimized.


\bibliography{OferBib}
%
\bibliographyfullrefs{OferBib}

\end{document}